\documentclass[nofootinbib]{revtex4}

\usepackage[usenames,dvipsnames]{color}
\usepackage{amsfonts,amssymb,amsmath,amsthm}
\usepackage{bm}
\usepackage[a4paper]{geometry}
\usepackage{pdflscape}
\usepackage{graphicx}
\usepackage{mathrsfs}

\definecolor{darkred}{rgb}{.8,0,0}

\definecolor{darkblue}{rgb}{0,0,.7}

\newcommand{\eps}{\varepsilon}

\newcommand{\scal}[1]{\langle #1 \rangle}
\newcommand{\rev}[1]{\widetilde{#1}}
\newcommand{\blade}[2]{{#1}_1 \wedge \ldots \wedge {#1}_{#2}}

\newcommand{\veclist}[2]{{#1}_1,\ldots,{#1}_{#2}}

\newcommand{\cf}[1]{\mathsf{#1}} 

\begin{document}

\title{Shape tensor and geometry of embedded manifolds}

\author{V\'{a}clav Zatloukal}

\email{zatlovac@gmail.com}

\homepage{http://www.zatlovac.eu}

\affiliation{\vspace{3mm}
Faculty of Nuclear Sciences and Physical Engineering, Czech Technical University in Prague, \\
B\v{r}ehov\'{a} 7, 115 19 Praha 1, Czech Republic \\
}

\begin{abstract}
We review the notion of shape tensor of an embedded manifold, which efficiently combines intrinsic and extrinsic geometry, and allows for intuitive understanding of some basic concepts of classical differential geometry, such as parallel transport, covariant differentiation, and curvature. We introduce shape-minimizing curves, i.e., curves between given two points that minimize integrated value of the shape tensor magnitude.
\end{abstract}

\maketitle

\section{Introduction}


In this article we are concerned with both intrinsic and extrinsic geometry of manifolds of arbitrary dimension embedded in an arbitrary-dimensional Euclidean space. The central object of this study, the \emph{shape tensor}, is introduced, and its basic properties are reviewed, in Sec.~\ref{sec:ShapeTensor}. The shape tensor is closely related to the second fundamental form of a surface, and its significance for the study of differentiable manifolds was emphasized by D. Hestenes in his work on geometric algebra and calculus \cite{Hestenes,Hestenes2011}. 

In Sec.~\ref{sec:ParalTr}, the parallel transport is defined as a process when infinitesimal rotations implemented by the shape tensor are used to maintain tangency of vectors transported along the manifold. In this way we reproduce the standard Levi-Civita transport of classical differential geometry. The covariant derivative and curvature are then defined in a standard way in Secs.~\ref{sec:CovDer} and \ref{sec:Curvature}. 
Importantly, the prescription for covariant derivative can be directly applied to both tangent and transverse vectors. As a consequence, the curvature naturally splits into two parts --- the intrinsic and the extrinsic one. The shape-tensor-based approach therefore automatically encodes both intrinsic and extrinsic geometry of an embedded manifold.

Frames and coordinates play little role in our discussion, however, we use them in Sec.~\ref{sec:FramesCoord} to make contact with the traditional treatment of differential geometry of surfaces, in particular, to recover the connection coefficients.

In Sec.~\ref{sec:ShapeMin} the concept of shape-minimizing curves is introduced. They are defined as curves connecting two given points, for which integrated value of the shape tensor magnitude is minimal. (This should be compared to the definition of geodesics as distance-minimizing curves.) 

Finally, we provide three examples in Sec.~\ref{sec:Examples} to illustrate the ease and clarity of calculations when employing the shape tensor and geometric algebra techniques. It is advisable to consult the corresponding parts of these examples after each section of the article.

Apart from Sec.~\ref{sec:MinShape}, this work is largely inspired by (and for the most part summarizes the results of) monographs \cite{Hestenes} and \cite{DoranLas} about the mathematical language of geometric algebra and calculus, and its use in pure and applied mathematics and in physics. Although we do not attempt to provide a thorough introduction into this formalism, we do recall the basic definitions in Appendix~\ref{sec:GAGC} in the scope necessary for understanding the main text.

\section{Shape tensor} \label{sec:ShapeTensor}

Let $\mathcal{M}$ be a smooth $n$-dimensional manifold embedded in the Euclidean space $\mathbb{R}^N$. At each point $x \in \mathcal{M}$, we identify the tangent space $T_x\mathcal{M}$, including a choice of its orientation, with a simple unit multivector of grade $n$, $I_\mathcal{M}(x)$. $I_\mathcal{M}$ is referred to as the \emph{pseudoscalar} of manifold $\mathcal{M}$. It is a smooth function of $x$ provided $\mathcal{M}$ is smooth. 
\footnote{Each tangent space possesses two unit pseudoscalars, which differ by a minus sign. Manifolds are orientable if $I_\mathcal{M}(x)$ is a single-valued function, and we shall assume this to be the case throughout the text.}
\footnote{The mapping $x \mapsto I_\mathcal{M}(x)$ generalizes the Gauss map that assigns to every point on $\mathcal{M}$ the corresponding normal vector, but which is uniquely defined only if the codimension of $T_x\mathcal{M}$ is one.}

At each point $x \in \mathcal{M}$, we define the \emph{shape tensor}
\footnote{The function defined by Eq.~\eqref{ShapeTensorDef} is referred to as the \emph{curl} of the manifold in Ref.~\cite{Hestenes}, whereas the term `shape tensor' is reserved for a more general object. Our use of the term `shape tensor' is in accordance with Ref.~\cite[Ch.\,6.5.3]{DoranLas}.} 
as the directional derivative
\begin{equation} \label{ShapeTensorDef}
S(a) 
= I_\mathcal{M}^{-1} \, a \cdot \partial I_\mathcal{M} ,
\end{equation}
where $a$ is a tangent vector at $x$. $S(a;x)$ is therefore a linear function of $a$, and a generic function of $x$. It is illuminating to choose a frame of orthonormal tangent vector fields to express the pseudoscalar as $I_\mathcal{M} = e_1 \ldots e_n$, and cast
\begin{align} \label{ShapeTensorFrame}
S(a)
&= e_n \ldots e_1 \sum_{k=1}^n e_1 \wedge \ldots \wedge e_{k-1} \wedge [ \mathsf{P}(a \cdot \partial e_k) + \mathsf{P}_\perp(a \cdot \partial e_k) ] \wedge e_{k+1} \ldots \wedge e_n
\nonumber\\
&= e_k \wedge \mathsf{P}_\perp(a \cdot \partial e_k)
\end{align}
Here, $\mathsf{P}$ is the operator of projection onto the tangent space, $\mathsf{P}_\perp$ is the projection onto the transverse space (the orthogonal complement of the tangent space in the ambient space $\mathbb{R}^N$), and we have use the fact that $a \cdot \partial e_k$ is perpendicular to $e_k$ due to the normalization condition $e_k^2=1$. In the last expression, the Einstein summation over the repeated index is implied.

\begin{figure}[h]
\includegraphics[scale=1]{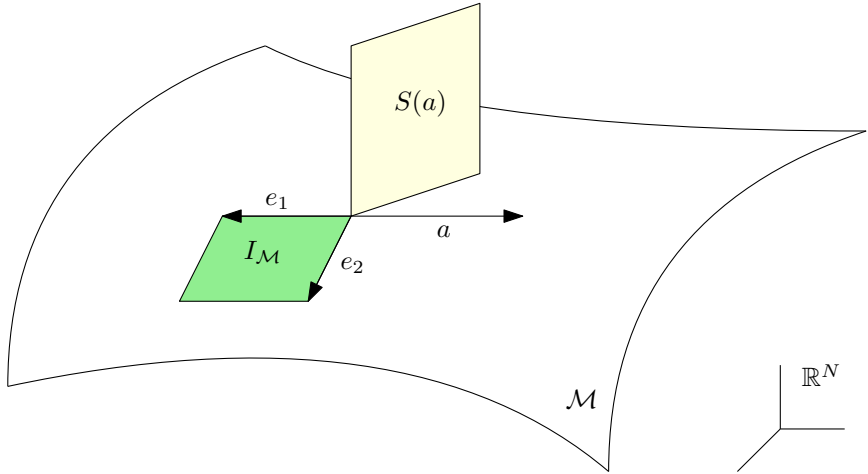}
\caption{The pseudoscalar $I_\mathcal{M}$ and the shape tensor $S(a)$ of an embedded manifold $\mathcal{M}$.} 
\label{fig:ShapeTensor}
\end{figure}
The representation~\eqref{ShapeTensorFrame} makes it obvious that $S(a)$ is a bivector (in general not simple) that consists of terms of the form (tangent vector)$\wedge$(transverse vector) (see Fig.~\ref{fig:ShapeTensor}).
The latter observation implies that $S(a)$ anticommutes
\footnote{This can be derived also by differentiating the equation $I_\mathcal{M}^{-1} I_\mathcal{M} = 1$, and taking into account that $I_\mathcal{M}^{-1} = (-1)^{n(n-1)/2} I_\mathcal{M}$.} 
with $I_\mathcal{M}$, 
\begin{equation} \label{ShapeTensorAnticomm}
S(a) I_\mathcal{M} = -I_\mathcal{M} S(a) ,
\end{equation}
and so we can write
\begin{equation} \label{ShapeTensorRot}
a \cdot \partial I_\mathcal{M}
= I_\mathcal{M} \times S(a) .
\end{equation}
We realize that the shape tensor can be interpreted as the angular velocity of the pseudoscalar $I_\mathcal{M}$ as it slides along the manifold, i.e., as the bivector field that implements infinitesimal rotations keeping $I_\mathcal{M}$ tangent.
\footnote{For hypersurfaces, i.e., submanifolds of dimension $N-1$, dotting Eq.~\eqref{ShapeTensorFrame} with a tangent vector $b$ and the unit normal $n$ gives $n \cdot \big( b \cdot S(a) \big) = n \cdot (a \cdot \partial b)$ --- the second fundamental form of $\mathcal{M}$ \cite[Ch.~11.4]{Frankel}.} 

Finally, let us remark that for a pair of tangent vector fields $a$ and $b$, the Lie bracket $[a,b] = a \cdot \partial \,b - b \cdot \partial \,a$ is again tangent, and Eq.~\eqref{ShapeTensorFrame} thus implies the following symmetry property of the shape tensor:
\begin{equation}
b \cdot S(a) = a \cdot S(b) .
\end{equation}

\section{Parallel transport} \label{sec:ParalTr}

\begin{figure}[h] 
\includegraphics[scale=1]{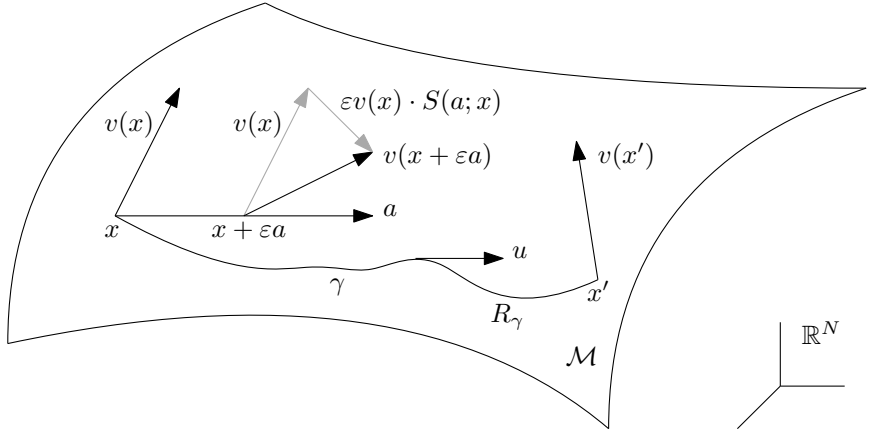}
\caption{Parallel transport of a vector $v(x)$ to a nearby point $x + \eps a$ as a composition of ambient translation and infinitesimal rotation by the bivector $S(a)$. Finite parallel transport along a curve $\gamma$ is performed by the rotor $R_\gamma$: $v(x')=R_\gamma(x')v(x)\rev{R}_\gamma(x')$.}
\label{fig:ParalTr}
\end{figure}

Parallel transport along the manifold $\mathcal{M}$ can be defined in an intuitive way using the shape tensor. Suppose $A(x)$ is a multivector at some point $x$ (which may contain tangent as well as transverse components), and define the parallel-transported multivector at a close point ${x+\eps a \in \mathcal{M}}$ by composing parallel transport in the ambient Euclidean space, and infinitesimal rotation by the bivector $S(a)$ (see Fig.~\ref{fig:ParalTr}):
\begin{equation} \label{ParalTrDef}
A(x + \eps a)
= A(x) + \eps A(x) \times S(a;x)
\approx e^{- \frac{\eps}{2} S(a)} A(x) e^{\frac{\eps}{2} S(a)} .
\end{equation}

According to this definition the pseudoscalar $I_\mathcal{M}$ is parallel transported along $\mathcal{M}$ (cf. Eq.~\eqref{ShapeTensorRot}). Since rotations commute with the geometric product (and hence with any other product composed from the geometric product, such as $``\cdot"$, $``\wedge"$ or $``\times"$), the parallel transport preserves tangency and transversality, as well as the inner product between vectors.
\footnote{For example, if $v(x)$ is tangent at $x$, i.e., $v(x) \wedge I_\mathcal{M}(x) = 0$, at a nearby point we find $$v(x + \eps a) \wedge I_\mathcal{M}(x+\eps a) = e^{- \frac{\eps}{2} S(a)} v(x) \wedge I_\mathcal{M}(x) e^{\frac{\eps}{2} S(a)} = 0 .$$}
\footnote{In an alternative definition of the parallel transport, tangent vectors are kept tangent by projection onto the tangent space of the manifold: $$v(x+\eps a) = v(x) \cdot e_j(x+\eps a) e_j(x+\eps a) .$$ Here, $e_j$'s form an orthonormal frame, and $v$ is assumed to be tangent at point $x$. Expanding the right-hand side we find $$\frac{1}{\eps} \big( v(x+\eps a) - v(x) \big) \approx v \cdot (a \cdot \partial e_j) e_j + v \cdot e_j a  \cdot \partial e_j  = - v \cdot e_k \mathsf{P}(a \cdot \partial e_k) + v \cdot e_k a \cdot \partial e_k = v \cdot \big(e_k \wedge \mathsf{P}_\perp(a \cdot \partial e_k) \big) ,$$ which agrees with Eq.~\eqref{ParalTrDef} by virtue of Formula~\eqref{ShapeTensorFrame}.} 

By Eq.~\eqref{ParalTrDef}, a multivector $A$ is parallel-transported along a curve $\gamma \subset \mathcal{M}$ (whose unit tangent vector we denote by $u$), if
\begin{equation}
u \cdot \partial A
= A \times S(u) .
\end{equation} 
In particular, if $u$ itself is parallel-transported, i.e., $u \cdot \partial u
= u \cdot S(u)$, then $\gamma$ is a geodesic.

In order to achieve a finite parallel transport we integrate the infinitesimal rotations of Eq.~\eqref{ParalTrDef} along $\gamma$ to obtain a finite rotor $R_\gamma$, which takes $A(x)$ at some point $x$ into a parallel multivector $A(x') = R_\gamma(x') A(x) \rev{R}_\gamma(x')$. It satisfies $R_\gamma(x + \eps u) = e^{-\frac{\eps}{2} S(u;x)} R_\gamma(x)$, $\forall x \in \gamma$, or in other words, the differential equation 
\begin{equation}
u \cdot \partial R_\gamma 
= -\frac{S(u)}{2} R_\gamma
\quad,\quad R_\gamma(x) = 1 .
\end{equation}
The solution of this equation can be expressed as a path-ordered exponential
\begin{equation}
R_\gamma(x')
= \mathcal{P}\exp \left[- \int_{x}^{x'} \frac{S(d\Gamma)}{2} \right] ,
\end{equation}
where $d\Gamma = u |d\Gamma|$ is the line element of the curve $\gamma$.

The parallel transport between $x$ and $x'$ typically depends upon the choice of path between the two points, since different paths encounter different bivectors $S(u)$ along the way. In explicit terms, $R_{\gamma}(x') \neq R_{\gamma'}(x')$ in general. 
\footnote{When we consider closed loops $\gamma$, the rotors $R_\gamma$ form the holonomy group of the manifold $\mathcal{M}$.}

\section{Covariant derivative} \label{sec:CovDer}

With the notion of parallel transport in hand, one defines the covariant derivative of a multivector field $A(x)$ in a direction $a$ as the difference between the result of the parallel transport of $A$ from a nearby point $x+\eps a$ to $x$, and the value of $A$ at $x$:
\begin{align} \label{CovDerDef}
a \cdot D A
&= \lim_{\eps \rightarrow 0} \frac{1}{\eps}
\left[ A(x+\eps a) + \eps A(x+\eps a) \times S(-a;x+\eps a) - A(x) \right]
\nonumber\\
&= a \cdot \partial A(x) - A(x) \times S(a;x) .
\end{align}
Since the parallel transport preserves tangency and transversality of objects it acts on, so does the covariant derivative. Comparison with Eq.~\eqref{ParalTrDef} shows that $A$ is parallel-transported in direction $a$ if and only if the covariant derivative $a \cdot D A$ vanishes.

Several properties of the covariant derivative follow easily. 

If $v$ is a tangent vector field, $v \cdot S(a)$ is transverse, and so the projection of Eq.~\eqref{CovDerDef} onto the tangent space yields
\footnote{Recall that the commutator product of a vector and a bivector reduces to the inner product: $v \times S(a) = v \cdot S(a)$.}
\begin{equation} \label{CovDerProjTan}
a \cdot D v
= \mathsf{P}(a \cdot \partial v) .
\end{equation} 
Likewise, for a transverse vector field $w$, the expression $w \cdot S(a)$ is a tangent vector, and we find
\begin{equation} \label{CovDerProjTr}
a \cdot D w
= \mathsf{P}_\perp(a \cdot \partial w) .
\end{equation}

The covariant Leibniz rule is a consequence of the Leibniz rule for ordinary derivatives, and identity~\eqref{GAGC:CommLeib}:
\begin{equation} \label{CovDerLeibniz}
a \cdot D (A B)
= a \cdot \partial (A B) - (A B) \times S(a)
= (a \cdot D A) B + A (a \cdot D B) .
\end{equation}

Mere rewriting of Eq.~\eqref{ShapeTensorRot} demonstrates that the pseudoscalar of manifold $\mathcal{M}$ is covariantly conserved, 
\begin{equation}
a \cdot D I_\mathcal{M} = 0 .
\end{equation}

Another important result concerns covariant derivative of the shape tensor. From its definition~\eqref{ShapeTensorDef}, and the anticommutation property~\eqref{ShapeTensorAnticomm}, we find
\begin{equation} \label{CovDerShapeTensor}
a \cdot D S(b) - b \cdot D S(a)
= a \cdot \partial S(b) - b \cdot \partial S(a) - 2 S(b) \times S(a)
= S([a,b]) ,
\end{equation}
where $a$ and $b$ are arbitrary tangent vector fields.

\section{Curvature} \label{sec:Curvature}

Curvature of a manifold is introduced as a measure of non-commutativity of covariant derivatives in different directions. Therefore, let $A$ be a `test' multivector field, and let us calculate
\begin{align}
( a \cdot D \, b \cdot D - b \cdot D \, a \cdot D ) A 
&= a \cdot D \big( b \cdot \partial A - A \times S(b) \big) - b \cdot D \big( a \cdot \partial A - A \times S(a) \big)
\nonumber\\
&= [a,b] \cdot \partial A + (A \times S(a)) \times S(b) - (A \times S(b)) \times S(a) - A \times S([a,b])
\nonumber\\
&= [a,b] \cdot D A + A \times (S(a) \times S(b)) .
\end{align}
We have used the definition of covariant derivative, Eq.~\eqref{CovDerDef}, the covariant Leibniz rule, Eq.~\eqref{CovDerLeibniz}, the identity~\eqref{CovDerShapeTensor}, and, finally, the Jacobi identity~\eqref{GAGC:JacobiId}. The final result features the quantity
\begin{equation} \label{CurvOmega}
\Omega(a \wedge b)
= S(a) \times S(b) ,
\end{equation}
which we call (in agreement with \cite[Ch.~5-1]{Hestenes}) the \emph{(total) curvature}. It is an $x$-dependent bivector-valued linear function of a tangent bivector argument.
\footnote{The last claim can be understood with a help of an orthonormal basis of the tangent space, since one can then write explicitly $$S(a) \times S(b) = \frac{1}{2} (a \wedge b) \cdot (e_j \wedge e_k) S(e_k) \times S(e_j) .$$} 
\footnote{Eq.~\eqref{CurvOmega} defines $\Omega$ only for simple bivectors $a \wedge b$, but we may linearly extend the definition to all tangent bivectors.}

\begin{figure}[h] 
\includegraphics[scale=1]{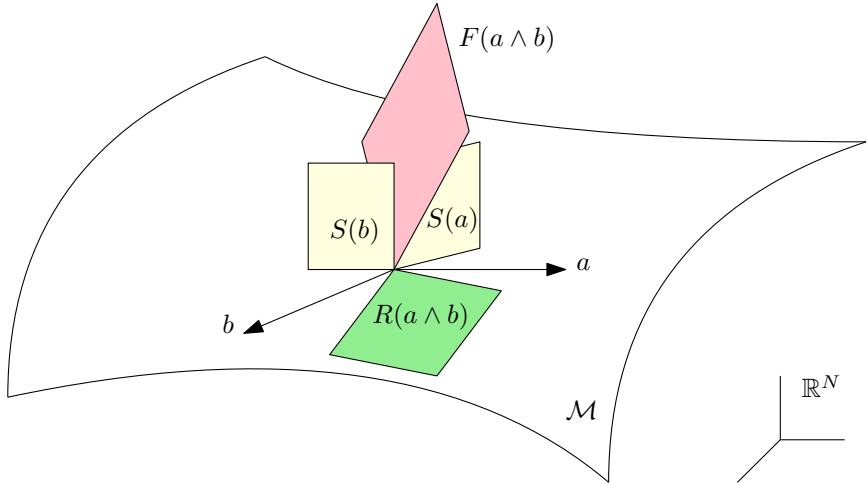}
\caption{Curvature is the commutator product of the shape tensor in two directions $a$ and $b$. It splits into the tangential (intrinsic) part $R(a \wedge b)$, and the transverse (extrinsic) part $F(a \wedge b)$.}
\label{fig:Curvature}
\end{figure}

The total curvature, being a commutator product of two bivectors of the form (tangent vector)$\wedge$(transverse vector), is a sum of a purely tangent bivector, and purely a transverse bivector.
\footnote{This becomes clear once we realize that for two bivectors $v_1 \wedge w_1$ and $v_2 \wedge w_2$, where $v_1$, $v_2$ are tangent and $w_1$, $w_2$ transverse, $$(v_1 \wedge w_1) \times (v_2 \wedge w_2) = - v_1 \cdot v_2 \, w_1 \wedge w_2 - w_1 \cdot w_2 \, v_1 \wedge v_2 .$$} 
The tangent, or intrinsic, part is commonly referred to as the \emph{Riemann tensor} 
\begin{equation}
R(a \wedge b)
= \mathsf{P}(S(a) \times S(b)) .
\end{equation}
The transverse, or extrinsic, part of the curvature is denoted by
\begin{equation}
F(a \wedge b)
= \mathsf{P}_{\perp}(S(a) \times S(b)) .
\end{equation}
[See the cartoon representation of Fig.~\ref{fig:Curvature}.]

Note that since the shape tensor lies partly in the tangent and partly in the transverse space, its scalar product with the curvature bivector always vanishes.

\section{Frames and coordinates} \label{sec:FramesCoord}

In this section we choose, smoothly at each point of the manifold, a basis $\{ e_j \}_{j=1}^n$ of the tangent space, and a basis $\{ e_b \}_{b=n+1}^{N}$ of the transverse space. The vector fields $\veclist{e}{n}$ form a tangent frame, while the fields $e_{n+1}, \ldots, e_N$ form a transverse frame. The respective reciprocal frames consist of the unique vector fields $\{ e^j \}_{j=1}^n$ and $\{ e^b \}_{b=n+1}^{N}$ that satisfy
\begin{equation}
e_j \cdot e^k = \delta_j^k
\quad,\quad
e_b \cdot e^c = \delta_b^c .
\end{equation}

With the basis decomposition of a generic vector $v = v^j e_j + v^b e_b$, where the components $v^j = v \cdot e^j$ and $v^b = v \cdot e^b$ are scalar functions, we may cast the covariant derivative as
\begin{equation}
a \cdot D v
= (a \cdot \partial v^j) e_j + v^j \Gamma(a)_j^k e_k
+ (a \cdot \partial v^b) e_b + v^b \Pi(a)_b^c e_c ,
\end{equation}
where
\begin{equation}
\Gamma(a)_j^k
= (a \cdot \partial e_j) \cdot e^k  
\quad,\quad
\Pi(a)_b^c
= ( a \cdot \partial e_b ) \cdot e^c 
\end{equation}
are the connection coefficients in the tangent and transverse space, respectively. [We have used the covariant Leibniz rule, and Eqs.~\eqref{CovDerProjTan} and \eqref{CovDerProjTr}.]

There is a preferred basis of the tangent space when local coordinates on $\mathcal{M}$ are given --- the coordinate frame $\{ \cf{e}_j \}_{j=1}^n$, with the property $[\cf{e}_j,\cf{e}_k] = 0$. The coefficients $\Gamma(a)_j^k$ are then the Christoffel symbols, which can be expressed in terms of the metric $g_{ij} = \cf{e}_i \cdot \cf{e}_j$, and its inverse $g^{ij} = \cf{e}^i \cdot \cf{e}^j$ in a standard way.

\begin{figure}[h]
\includegraphics[scale=1]{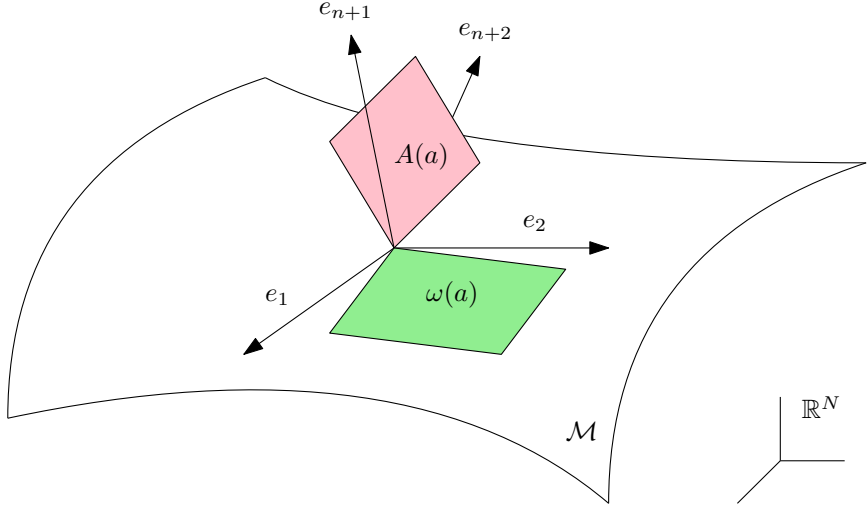}
\caption{Tangent frame $\{\veclist{e}{n}\}$, and transverse frame $\{e_{n+1},\ldots,e_N\}$. For frames with constant inner products, the covariant derivative is determined by the connection bivectors $\omega(a)$ (tangent), and $A(a)$ (transverse).}
\label{fig:Frames}
\end{figure}

Another option to specify frames is to demand that the scalar products $e_j \cdot e_k$, $e_b \cdot e_c$ be constant in $x$ for all $j,k,b,c$. We assume this to be the case until the end of this section. The covariant derivative can now be cast in a somewhat more transparent form
\begin{equation} \label{CovDerONB}
a \cdot D v
= (a \cdot \partial v^j) e_j + v \cdot \omega(a)
+ (a \cdot \partial v^b) e_b + v \cdot A(a) ,
\end{equation}
where
\begin{equation}
\omega(a)
= \frac{1}{2} e^j \wedge (a \cdot D e_j) 
\quad,\quad
A(a) = \frac{1}{2} e^b \wedge (a \cdot D e_b)
\end{equation}
are tangent- and transverse-space bivectors, respectively (see Fig.~\ref{fig:Frames}). They are referred to as the \emph{connection bivectors}.

Working out now the commutator of covariant derivatives, we obtain an expression for the intrinsic and extrinsic curvatures in terms of $\omega$ and $A$:
\begin{align} \label{RFcov}
R(a \wedge b)
&= a \cdot D \omega(b) - b \cdot D \omega(a)
+ \omega(a) \times \omega(b) - \omega([a,b]) ,
\nonumber\\
F(a \wedge b)
&= a \cdot D A(b) - b \cdot D A(a)
+ A(a) \times A(b) - A([a,b]) .
\end{align}

Yet another representation is derived by utilizing the bivector bases
$e_J = e_{j_1} \wedge e_{j_2}$, $e_B = e_{b_1} \wedge e_{b_2}$, and their reciprocals $e^J = e^{j_2} \wedge e^{j_1}$, $e^B = e^{b_2} \wedge e^{b_1}$ to expand
\begin{equation}
\omega(a)
= \omega(a)^J e_J
\quad,\quad
A(a)
= A(a)^B e_B
\end{equation}
where
$\omega(a)^J = \omega(a) \cdot e^J$ and $A(a)^B = A(a) \cdot e^B$.
[Summations over repeated multi-indices $J=(j_1,j_2)$, $j_1<j_2$, and $B = (b_1,b_2)$, $b_1<b_2$, are implied.] 
Note that for $v = e_j$ Eq.~\eqref{CovDerONB} simplifies to $a \cdot D e_j = e_j \cdot \omega(a)$, and the covariant Leibniz rule then gives $a \cdot D e_J = e_J \times \omega(a)$. Analogously, $a \cdot D e_B = e_B \times A(a)$.
We thus find
\begin{align}
a \cdot D \omega(b)
&= \big( a \cdot \partial \omega(b)^J \big) e_J
+ \omega(b) \times \omega(a) ,
\nonumber\\
a \cdot D A(b)
&= \big( a \cdot \partial A(b)^B \big) e_B
+A(b) \times A(a) ,
\end{align}
and finally arrive at
\begin{align} \label{RFComp}
R(a \wedge b)
&= \big( a \cdot \partial \omega(b)^J - b \cdot \partial \omega(a)^J \big) e_J
+ \omega(b) \times \omega(a) - \omega([a,b]) ,
\nonumber\\
F(a \wedge b)
&= \big( a \cdot \partial A(b)^B - b \cdot \partial A(a)^B \big) e_B
+ A(b) \times A(a) - A([a,b]) .
\end{align}

\section{Shape-minimizing curves} \label{sec:MinShape}
\label{sec:ShapeMin}

In analogy with geodesics, i.e., curves that minimize distance between two given points $x_1$ and $x_2$, let us study curves that minimize the functional
\begin{equation}
\Sigma[\gamma]
= \int_\gamma |S(d\Gamma)| ,
\end{equation}
where $d\Gamma$ is the oriented line element of a curve $\gamma$, which can be written as a product $d\Gamma = u |d\Gamma|$ of the unoriented line element $|d\Gamma|$, and the unit tangent vector $u$. We refer to curves that minimize the functional $\Sigma[\gamma]$ as \emph{shape-minimizing curves}.

Let $\gamma' = \{ x + \eps a(x) \,|\, x \in \gamma\}$ be a variation of $\gamma$, governed by a vector field $a$, that fixes the endpoints: $a(x_1) = a(x_2) = 0$. The line element and the shape tensor are varied as follows:
\begin{equation}
d\Gamma' 
= d\Gamma + \eps \,d\Gamma \cdot \partial \, a 
= |d\Gamma| \big( u + \eps [ u , a ] + \eps a \cdot \partial u \big)
\end{equation}
and
\begin{equation}
S(d\Gamma';x')
\approx S(d\Gamma) + \eps |d\Gamma| \big( S([u,a]) + a \cdot \partial S(u) \big)
\end{equation}
Taking into account Eq.~\eqref{CovDerShapeTensor}, we can further rearrange
\begin{equation}
S(d\Gamma';x')
\approx S(d\Gamma) + \eps \big( d\Gamma \cdot \partial S(a) + 2 S(d\Gamma) \times S(a) \big) .
\end{equation}

Variation of the functional $\Sigma[\gamma]$ is carried out conveniently with a help of Formula~\eqref{GAGC:MagnitudeExpansion},
\begin{equation}
\delta \Sigma[\gamma]
\approx \eps \int_\gamma \frac{\rev{S}(d\Gamma)}{|S(d\Gamma)|} \cdot \big( d\Gamma \cdot \partial S(a) + 2 S(d\Gamma) \times S(a) \big)
= - \eps \int_\gamma \rev{S}(a) \cdot \left[ d\Gamma \cdot \partial \frac{S(u)}{|S(u)|} \right] ,
\end{equation}
where we have made integration by parts, and noticed that $\rev{S}(d\Gamma) \cdot (S(d\Gamma) \times S(a)) = 0$ since the scalar product between the curvature and the shape bivectors always vanishes. (They belong to different bivector subspaces.)
Extremal curves satisfy the differential equation
\begin{equation} \label{ShapeMinDiffEq}
\rev{S}(a) \cdot \left[ u \cdot \partial \frac{S(u)}{|S(u)|} \right]
= 0
\end{equation}
for all tangent vectors $a$.

The derivation, and the final equation~\eqref{ShapeMinDiffEq}, only make sense under the assumption $|S(u)| > 0$. The expression $\rev{S}(a) \cdot S(b)$ can be understood as an alternative metric on the manifold $\mathcal{M}$, which, however, may well be degenerate.
\footnote{Note that for a plane (for example), the shape tensor identically vanishes, and so all curves have the same value of $\Sigma[\gamma]$, which is equal to 0.}

The partial derivative $u \cdot \partial$ in Eq.~\eqref{ShapeMinDiffEq} can be replaced by a covariant derivative $u \cdot D$, as these two only differ by a term proportional to $S(u) \times S(u) = 0$. Then, it is clear that the bivector in the square bracket is composed of terms of the form (tangent vector)$\wedge$(transverse vector). Therefore, if the space of such bivectors is spanned by the shape bivectors $S(a)$, which will be the case in Example~\ref{sec:ExEllipsoid}, Eq.~\eqref{ShapeMinDiffEq} is equivalent to
\begin{equation} \label{ShapeMinDiffEq2}
u \cdot \partial \frac{S(u)}{|S(u)|}
= 0 .
\end{equation}

\section{Examples} \label{sec:Examples}

\subsection{Curve} \label{sec:ExCurve}

\begin{figure}[h] 
\includegraphics[scale=1]{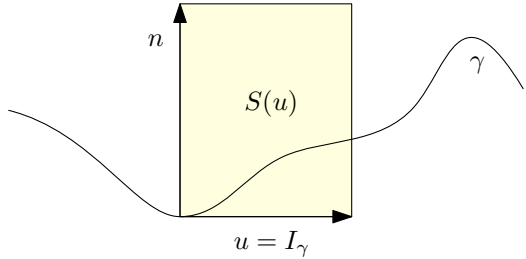}
\caption{Shape tensor of a curve is the bivector $S(u)=u n$, where $n$ is the normal and $u$ is the unit tangent vector.}
\label{fig:Curve}
\end{figure}

Let $\mathcal{\gamma}$ be a one-dimensional curve (depicted in Fig.~\ref{fig:Curve}), in which case the pseudoscalar $I_\mathcal{\gamma} = u$ is simply the unit tangent vector.
The shape tensor is the simple bivector
\begin{equation}
S(u) 
= u \, u \cdot \partial u
= u \,n ,
\end{equation}
where $n$ is the normal vector of $\gamma$.
\footnote{For a curve $\mathcal{\gamma} = \{ x(\tau) \,|\, \tau \in (\tau_1,\tau_2) \}$, parametrized by its arc-length $\tau$, the unit tangent is the first derivative $u = \frac{d x}{d \tau}$, and the normal vector is the second derivative $n = \frac{d u}{d \tau} = \frac{d^2 x}{d\tau^2}$.}
Curvature of the curve $\gamma$ is the magnitude of $S(u)$,
\begin{equation}
|S(u)|
= \sqrt{(n \,u) \cdot (u \,n)}
= |n| ,
\end{equation}
which coincides with the magnitude of the normal.

Parallel transport of tangent vectors is trivial, since there is only a single tangent vector field up to scalar multiplication, and it is parallel as long as it does not change the magnitude. For transverse vectors, the parallel transport defined in terms of the shape tensor can be used to construct a parallel frame that consists of the unit tangent, and unit normals that rotate the least possible amount about the tangent when transported along the curve $\gamma$. In the theory of relativity this transport is referred to as \emph{Fermi-Walker}.

\subsection{Sphere} \label{sec:ExSphere}

\begin{figure}[h] 
\includegraphics[scale=1]{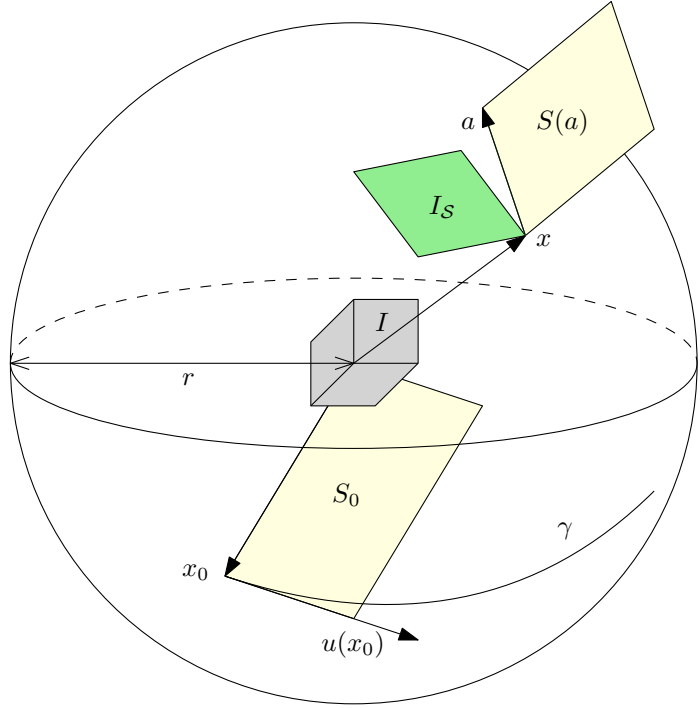}
\caption{Sphere of radius $r$ with pseudoscalar $I_\mathcal{S}$, embedded in a Euclidean space with pseudoscalar $I$. A geodesic $\gamma$ is constructed by rotating a point $x_0 \in \mathcal{S}$ in the plane specified by the bivector $S_0 = \frac{1}{r^2} x_0 u(x_0) = S(u(x_0);x_0)$.}
\label{fig:Sphere}
\end{figure}

The $n$-dimensional sphere of radius $r$ is the set $\mathcal{S} = 
\{ x \in \mathbb{R}^{n+1} \,|\, x^2=r^2 \}$.
The pseudoscalar of $\mathcal{S}$ at a point $x \in \mathcal{S}$ is given by
\begin{equation}
I_\mathcal{S}(x) 
= I \frac{x}{r} ,
\end{equation}
where $I$ is the unit pseudoscalar of the ambient space $\mathbb{R}^{n+1}$ (see Fig.~\ref{fig:Sphere}).

The shape tensor of the sphere reads
\begin{equation}
S(a) 
= \frac{1}{r^2} x I^{-1} \, a \cdot \partial (I x)
= \frac{x a}{r^2} ,
\end{equation}
and it generates the curvature
\begin{equation}
\Omega(a \wedge b)
= S(a) \times S(b)
= - \frac{a \wedge b}{r^2} ,
\end{equation}
which has only the intrinsic part: $\Omega(a \wedge b) = R(a \wedge b)$.
\footnote{Vanishing of the extrinsic part of the curvature $F(a \wedge b)$ was anticipated as there is no ``room" for a non-zero bivector in the one-dimensional transverse space.}
We observe that since the expression
\begin{equation}
\rev{S}(a) \cdot S(b)
= \frac{a \cdot b}{r^2}
\end{equation}
differs from the metric only by a constant scalar factor $1/r^2$, the shape-minimizing curves on a sphere coincide with the distance-minimizing curves --- the geodesics.

A geodesic $\gamma$  (with unit tangent vector $u$) fulfils the equation
\begin{equation}
u \cdot \partial u
= u \cdot S(u)
= \frac{1}{r^2} u \cdot (x \wedge u) 
= - \frac{x}{r^2} .
\end{equation}
This equation implies a conservation law for the shape tensor,
\begin{equation}
u \cdot \partial S(u) 
= \frac{1}{r^2} u \cdot \partial (x u)
= 0 ,
\end{equation}
meaning that 
\begin{equation}
\frac{1}{r^2} x u
= \frac{1}{r^2} x_0 u(x_0)
= S_0
\end{equation}
for some constant bivector $S_0$, which defines the initial position and direction of the motion.
Introducing an arc-length parametrization of $\gamma$, we now have to integrate only the first-order differential equation
\begin{equation}
u = \frac{d x}{d \tau}
= x S_0 ,
\end{equation}
which is straightforward:
\begin{equation}
x(\tau) = x_0 e^{\tau S_0}
= e^{-\frac{\tau}{2} S_0} x_0 e^{\frac{\tau}{2} S_0} .
\end{equation}
We have found that geodesic on a sphere is an arc that lies in the plane spanned by the bivector $S_0$.

\subsection{Ellipsoid} \label{sec:ExEllipsoid}

In this example we consider an $n$-dimensional ellipsoid
\begin{equation}
\mathcal{E}
= \{ x \in \mathbb{R}^{n+1} \,|\, \phi(x) = 1 \}
\quad,\quad
\phi(x) = x \cdot A(x) ,
\end{equation}
where $A$ is a constant symmetric positive-definite linear mapping.
The pseudoscalar of $\mathcal{E}$ is $I_\mathcal{E} = I n(x)$, with the unit normal
\begin{equation}
n(x)
= \frac{\partial_x \phi}{|\partial_x \phi|}
= \frac{A(x)}{|A(x)|} .
\end{equation}
The shape tensor reads
\footnote{Any linear map $A$ can be naturally extended to act on arbitrary multivectors by defining $A(\blade{a}{r}) = A(a_1) \wedge \ldots \wedge A(a_r)$ (see the definition of \emph{outermorphism} in \cite[Ch.~3-1]{Hestenes}).} 
\begin{equation}
S(a)
= n \wedge (a \cdot \partial n)
= \frac{A(x \wedge a)}{|A(x)|^2} ,
\end{equation}
and the curvature is
\begin{equation}
S(a) \times S(b)
= -\frac{1}{|A(x)|^2}
\left[ A(a \wedge b) - n \wedge \big( n \cdot A(a \wedge b) \big) \right] .
\end{equation}
It has only intrinsic part, as can be shown explicitly by taking inner product with the normal $n$.

It follows from non-degeneracy of the linear map $A$ that $S(a) \neq 0$ for all non-zero tangent vectors $a$, and hence the shape-minimizing curves $\gamma$ are solutions of Eq.~\eqref{ShapeMinDiffEq2}, which for the ellipsoid reads
\begin{equation} \label{ShapeMinEllipsoid}
u \cdot \partial \frac{A(x \wedge u)}{|A(x \wedge u)|}
= 0 ,
\end{equation}
where $u$ is the unit tangent of $\gamma$.
To solve, choose parametrization $\tau$ of $\gamma = \{x(\tau)\}$ such that
\begin{equation}
\frac{d x}{d \tau}
\equiv \dot{x}
= \frac{u}{|A(x \wedge u)|} ,
\end{equation}
and integrate \eqref{ShapeMinEllipsoid} to get
\begin{equation}
A(x \wedge \dot{x}) = A(B_0) ,
\end{equation}
where $B_0$ is a bivector constant of integration.
Now apply $A^{-1/2}$,
and take inner product with $A^{1/2}(x)$ to obtain
\footnote{Using the fact that $x \in \mathcal{E}$, and $\dot{x}$ is perpendicular to $A(x)$ we calculate $$A^{1/2}(x) \cdot \big( A^{1/2}(x) \wedge A^{1/2}(\dot{x}) \big)
= x \cdot A(x) \, A^{1/2}(\dot{x}) - A(x) \cdot \dot{x} \, A^{1/2}(x) = A^{1/2}(\dot{x}).$$}
\begin{equation}
\frac{d}{d\tau} A^{1/2}(x) = A^{1/2}(x) \cdot A^{1/2}(B_0) .
\end{equation}
This equation is readily solved (see Eqs.~\eqref{GAGC:DiffEq} and \eqref{GAGC:DiffEqSolution}), and we obtain a shape-minimizing curve originating at point $x(0) = x_0$ in direction $u(0) = u_0$:
\begin{equation} \label{ShapeCurveEllipsoid}
x(\tau)
= A^{-1/2}\big( R(\tau) A^{1/2}(x_0) \rev{R}(\tau) \big)
\quad,\quad
R(\tau) = e^{-\frac{\tau}{2} A^{1/2}(B_0)} 
\quad,\quad
B_0 = \frac{x_0 \wedge u_0}{|A(x_0 \wedge u_0)|} .
\end{equation}

The shape-minimizing curve $x(\tau)$ does not leave the two-dimensional plane defined by the bivector $B_0$.
\footnote{Compare the relatively simple and explicit result~\eqref{ShapeCurveEllipsoid} with the discussion of geodesics on an ellipsoid \cite[Ch.\,3.5]{Klingenberg}.}

\section{Conclusion}

In this article we reintroduced some basic concepts of differential geometry (parallel transport, covariant derivative, curvature) in the case of embedded manifolds. We did so with a help of the shape tensor --- an object, which encodes intrinsic as well as extrinsic geometry of the manifold. 

In analogy with geodesics, we introduced shape-minimizing curves as extremals of the functional that measures the `shape'-distance, i.e., the integrated value of the shape tensor magnitude accumulated along the path. They have been found explicitly on a generic ellipsoid.

To phrase the presented ideas in a more abstract setting, one may consider an $N$-dimensional vector bundle over an $n$-dimensional base manifold $\mathcal{M}$, together with a map relating the tangent spaces $T_x\mathcal{M}$ with certain $n$-dimensional subspaces of the fibres. The complements of these subspaces are then the transverse spaces. Over each fibre we can introduce a geometric algebra, and the shape tensor can be regraded as a one-form with values in its bivector part. Further details shall be discussed in a separate article.


\appendix
\section{Elements of geometric algebra and calculus} 
\label{sec:GAGC}

In this appendix we introduce the machinery of geometric algebra and calculus in the extend necessary for understanding of expressions and manipulations used in the main text.
We mostly refrain from presenting proofs as these can be found in the standard textbooks on geometric algebra \cite{Hestenes,DoranLas}.

Let $V$ be an $N$-dimensional real vector space (in our case, $V \simeq \mathbb{R}^N$ is the tangent space of the ambient embedding space $\mathbb{R}^N$). For any two vectors $a$ and $b$ we define the \emph{geometric} (or \emph{Clifford}) product $a b$, which is associative, $a (b c) = (a b) c$, distributive, $a (b + c) = a b + a c$, and such that $a^2$ is a non-negative scalar, which vanishes only for the zero vector.
\footnote{To study spaces with mixed signature one can allow also negative or null squares  \cite[Ch.~1-5]{Hestenes}. We will, however, consider only positive signature, for simplicity.}
Splitting of the geometric product into the symmetric and antisymmetric part defines the \emph{inner} and \emph{outer} product, respectively:
\begin{equation}
a \cdot b
= \frac{1}{2}(a b + b a)
= b \cdot a
\quad,\quad
a \wedge b
= \frac{1}{2}(a b - b a)
= - b \wedge a .
\end{equation}

\emph{Multivectors} arise from products of multiple vectors. They span the entire \emph{geometric algebra} over $V$, $\mathcal{G}(V)$. A generic multivector $A = \sum_{r=0}^N A_r$ is a sum of terms with definite \emph{grade} $r$. A multivector has grade $r$, if it can be written as an outer product of $r$ vectors, $\blade{a}{r}$, or a sum of such products.
\footnote{If one term is enough, i.e., $A = \blade{a}{r}$, then $A$ is called a \emph{simple} (or decomposable) multivector of grade $r$, or, an $r$-\emph{blade}.}
The inner and outer product are extended to multivectors by the formulas
\footnote{Geometrically, an $r$-blade $A_r = \blade{a}{r}$ can be pictured as a parallelogram spanned by the vectors $\veclist{a}{r}$, the inner product lowers its dimension by one, ``squashing" the parallelogram along the direction $a$, while the outer product raises the dimension by one by expansion along $a$.}
\begin{equation}
a \cdot A_r 
= \frac{1}{2}\big(a A_r - (-1)^r A_r a \big)
= (-1)^{r-1} A_r \cdot a
\quad,\quad
a \wedge A_r 
= \frac{1}{2}\big(a A_r + (-1)^r A_r a \big)
= (-1)^r A_r \wedge a .
\end{equation}

Due to the antisymmetry of the outer product of vectors, there is a highest-grade element $I$ (of grade $N$) called the \emph{pseudoscalar} of the geometric algebra $\mathcal{G}(V)$, which is unique up to a scalar multiplication. The total dimension of $\mathcal{G}(V)$ is the sum of dimensions of individual equi-grade subspaces, and is equal to $\sum_{r=0}^N {N \choose r} = 2^N$.

The \emph{magnitude} of a generic multivector $A$ is defined by
\begin{equation} \label{GAGC:Magnitude}
|A|
= \sqrt{\scal{\rev{A} A}} ,
\end{equation}
where $\scal{\,.\,}$ denotes the scalar (or grade-$0$) part
\footnote{This corresponds to the operation of normalized trace in a matrix representation of the geometric algebra (cf., Dirac $\gamma$ matrices).}
of a multivector, and $\,\rev{.}\,$ is the \emph{reversion} operation, which reverses the order of vectors in a geometric product: $\rev{a\ldots b} = b \ldots a$.
For an $r$-blade $A_r$, the magnitude $|A_r|$ is equal to the volume of the corresponding parallelogram, and $A_r$ has an inverse $A_r^{-1} = \rev{A}_r/|A_r|^2$, where $\rev{A}_r = (-1)^{r(r-1)/2} A_r$.

The \emph{commutator} product between multivectors is defined by
\begin{equation}
A \times B
= \frac{1}{2}(A B - B A) .
\end{equation}
It satisfies the Leibniz rule
\begin{equation} \label{GAGC:CommLeib}
(A B) \times C
= A (B \times C) + (A \times C) B ,
\end{equation}
and the Jacobi identity
\begin{equation} \label{GAGC:JacobiId}
A \times (B \times C)
= (A \times B) \times C - (A \times C) \times B .
\end{equation}
If $B$ is a bivector (a multivector of grade $2$), and $A$ has grade $r$, then $A \times B$ has also grade $r$. Thereupon, the set of all bivectors is closed under the commutator product.

At this point, let us note that we adopt the convention according to which the $\,\cdot\,$, $\wedge$ and $\times$ products have always priority before the geometric product.

The following formulas can be useful for simplification of inner and commutator products:
\begin{align}
a \cdot (\blade{a}{r})
&= \sum_{j=1}^r (-1)^{j-1} a \cdot a_j \, a_1 \wedge \ldots \wedge a_{j-1} \wedge a_{j+1} \wedge \ldots \wedge a_r
\nonumber\\
B \times (\blade{a}{r})
&= \sum_{j=1}^r a_1 \wedge \ldots \wedge a_{j-1} \wedge (B \cdot a_j) \wedge a_{j+1} \wedge \ldots \wedge a_r ,
\end{align}
where $a$ is a vector, and $B$ is a bivector. Also note that for two vectors, $a \times b = a \wedge b$, and for a vector and a bivector, $a \times B = a \cdot B$.

Orthogonal transformations (rotations) on $V$ can be represented in the form
\begin{equation} \label{GAGC:Rotation}
a \mapsto 
a' = R a \rev{R} ,
\end{equation}
where the \emph{rotor} $R = e^{-B/2}$ is an exponential of a bivector (defined by power series with powers taken in the sense of geometric product), and $\rev{R} = e^{B/2}$ is the inverse rotor.
\footnote{If $B = b_1 \wedge b_2$ is a simple bivector, then Eq.~\eqref{GAGC:Rotation} performs rotation by angle $|B|$ in the plane spanned by the vectors $b_1$ and $b_2$,  in the sense ``from $b_1$ towards $b_2$".}
\footnote{When $a \wedge B = 0$ (i.e., $a$ anticommutes with $B$), which is the case, in particular, if $B$ is simple and $a$ lies in the plane defined by $B$, then $R a \rev{R} = R^2 a = a \rev{R}^2$.}
Since $R \rev{R} = 1$, generic multivectors are rotated in a completely analogous way:
\begin{equation}
A \mapsto
A' = R A \rev{R} .
\end{equation}
For infinitesimal rotations we can approximate
\begin{equation}
A'
= e^{-\eps \frac{B}{2}} A\, e^{\eps \frac{B}{2}} 
\approx A + \eps A \times B ,
\end{equation}
which reduces to $a' \approx a + \eps a \cdot B$ for vectors.
\footnote{The expression $a \cdot B$ can be viewed as a linear map on the vector space $V$, which is antisymmetric: $b \cdot (a \cdot B) = (b \wedge a) \cdot B = -a \cdot (b \cdot B)$. In turn, any antisymmetric linear map can be represented as $a \cdot B$ for some bivector $B$.}

Let $\{e_1, \ldots, e_n\}$ be a set of orthogonal vectors. The linear subspace $\mathcal{E}_n = {\rm span}\{\veclist{e}{n}\}$ can be identified with simple unit multivector $I_n = e_1 \ldots e_n = \blade{e}{n}$ in the sense that any $a \in \mathcal{E}_n$ satisfies $a \wedge I_n = 0$, and vice versa. The orthogonal projection $\mathsf{P}$ of a vector $a \in V$ onto $\mathcal{E}_n$ can be expressed in terms of $I_n$ as
\begin{equation}
\mathsf{P}(a)
= a \cdot I_n I_n^{-1}
= \sum_{j=1}^n a \cdot e_j e_j .
\end{equation}
The projection $\mathsf{P}_\perp$ onto the orthogonal complement of $\mathcal{E}_n$ in $V$ is then the difference
\begin{equation}
\mathsf{P}_\perp(a)
= a - a \cdot I_n I_n^{-1}
= (a I_n - a \cdot I_n) I_n^{-1}
= a \wedge I_n I_n^{-1} .
\end{equation}

In geometric calculus \cite{Hestenes}, the derivative of a multivector-valued function $F(x)$, $x \in \mathbb{R}^N$, in direction of a vector $a$ is defined by the usual limit
\begin{equation}
a \cdot \partial F(x)
= \lim_{\eps \rightarrow 0} \frac{F(x+\eps a) - F(x)}{\eps} .
\end{equation}
The Leibniz rule holds for the geometric product of functions,
\begin{equation}
a \cdot \partial (F G)
= (a \cdot \partial F) G + F a \cdot \partial G ,
\end{equation}
and hence also for the other, derived, products $\,\cdot\,$, $\wedge$ and $\times$.

We conclude with two useful observations.

First, for any grade-$n$ multivectors $A$ and $B$, $A \neq 0$,
\begin{equation} \label{GAGC:MagnitudeExpansion}
|A + \eps B| - |A| 
\approx \eps \frac{\rev{A} \cdot B}{|A|}
\end{equation}
holds up to the first order in $\eps$, as follow from the definition of magnitude, Eq.~\eqref{GAGC:Magnitude}.
\footnote{It is also possible to take advantage of the results of \cite[Ch.~2-2]{Hestenes} on multivector derivatives, namely, of Eq.~(2.32).}

Second, consider the differential equation ($x \in \mathbb{R}^N$, $\tau \in \mathbb{R}$)
\begin{equation} \label{GAGC:DiffEq}
\frac{d x}{d \tau}
= x \cdot B_0 
\quad,\quad
x(0) = x_0 ,
\end{equation}
where $B_0$ is a constant bivector.
Its solution is the curve
\begin{equation} \label{GAGC:DiffEqSolution}
x(\tau)
= e^{-\frac{\tau}{2} B_0} x_0 e^{\frac{\tau}{2} B_0} .
\end{equation}



\begin{thebibliography}{50}

\bibitem{Hestenes}
D. Hestenes and G. Sobczyk, {\it Clifford Algebra to Geometric Calculus}, Springer (1987).

\bibitem{Hestenes2011}
D. Hestenes, {\it The Shape of Differential Geometry in Geometric Calculus}, In: L. Dorst and J. Lasenby (eds.), {\it Guide to Geometric Algebra in Practice}, Springer, London (2011).


\bibitem{DoranLas}
C. Doran and A. Lasenby, {\it Geometric Algebra for Physicists}, Cambridge Univ. Press (2007).

\bibitem{Frankel}
T. Frankel, {\it The Geometry of Physics: An Introduction}, 2nd. Ed., Cambridge Univ. Press (2004).

\bibitem{Klingenberg}
W. Klingenberg, {\it Riemannian Geometry}, W. de Gruyter (1982).

\end{thebibliography}
\end{document}